\documentclass[12pt]{article}

\begin{document}                                                     


\def\1{{\bf 1}}
\def\Z{{\bf Z}}
\def\ee{\end{equation}}
\def\be{\begin{equation}}
\def\l{\label}
\def\D{{\cal D}}
\def\sin{{\rm sin}}
\def\cos{{\rm cos}}
\def\f{{\bf \Phi}}
\def\v{\varphi}
\def\O{\bf {\cal O}}
\def\C{\bf C}
\def\CP{\bf CP}
\def\e{\rm e}
\def\0{\nonumber}
\def\eea{\end{eqnarray}}
\def\bea{\begin{eqnarray}}
\def\Tr{\rm Tr}
\def\IR{\bf R}
\def\ZZ{\bf Z}
\def\T{\tau}
\def\ep{\epsilon}



\hfill ULB-TH/03-04
\vspace{.5 cm}


\def\title#1{\centerline{\huge{#1}}}
\def\author#1{\centerline{\large{#1}}}
\def\address#1{\centerline{\it #1}}
\def\ack{{\bf Acknowledgments}$\quad$}
\def\Bibliography{\begin{thebibliography}}
\def\endbib{\end{thebibliography}}


\title{On Type II strings in exact superconformal 
}\title{
non-constant RR backgrounds}
\author{Giulio Bonelli\footnote{e-mail address: gbonelli@ulb.ac.be}}
\address{
Physique Theorique et Mathematique - Universite Libre de Bruxelles}
\address{
Campus Plaine C.P. 231; B 1050 Bruxelles, Belgium}

\begin{abstract}
An explicitly exact superconformal description is provided to some classes of 
Type II string theories in non constant RR backgrounds.
This is done by applying the manifest (2,2) approach of Berkovits and Maldacena
to Type II strings and by studying the condition of exact conformal
invariance of certain supersymmetric backgrounds.
We find a new set of exact type IIA strings with non constant
RR 2-form and 4-form curvatures
and for type IIB with non constant 3-form curvature.
\end{abstract}

\tableofcontents

\section{Introduction}

The study of string theory in curved backgrounds
is crucial to understand which are its truly allowed vacua
and to get new insights about its phase structure.
Generally, the conformal invariance of a background -- imposed mainly
because of our actual inability in quantizing the non critical string 
$\sigma$-model coupled with the 
world-sheet Liouville field -- is analyzed order by order in $\alpha'$ and 
there are very few handles on exactly decoupled string theories.

It appeared recently an interesting class of pp-wave solutions of type IIB 
\cite{mama}, generalizing the one studied in \cite{metsaev}, with non constant 
RR fluxes.
These string theories have been shown to admit a superconformal
formulation in \cite{bema} and have been shown to be exact (finite)
in \cite{bema,ruts}.
These solutions have been studied in the $SU(4)\times U(1)$
formalism \cite{GS}.
Within this framework the type IIB GS action on a flat background 
is written in terms of four complex chiral (2,2) superfields 
$X^{+\bar l}$, 
where $l=1,\dots,4$, as
\be
S_0^{IIB}=\frac{1}{2}\int d^2z d^4\kappa X^{+\bar l}X^{-l}
\l{22GSB}\ee
where
$D_-X^{+\bar l}=0$, $\bar D_-X^{+\bar l}=0$ and 
$D_+X^{- l}=0$, $\bar D_+X^{- l}=0$.
Here $D_\pm$ and $\bar D_\pm$ are the usual superderivatives.
The only subgroup of the original $SO(8)$ symmetry which remains manifest is 
a $SU(4)\times U(1)$ one.

Upon improving and completing 
the theory with a proper set of free chiral and semi-chiral
multiplets, the theory can be shown to admit an equivalent explicitly
superconformal description.
In particular, there are two kind of extensions to curved backgrounds
with RR fields, which preserve respectively (1,1) and (2,2) supersymmetry.
These extensions are quite special since they keep the exact superconformal
symmetry. This means that once the superconformal invariance is
established at one loop, than it is implied also at higher loops
just because of algebraic (in the superspace) reasons.
All this is well explained in \cite{bema} and will not be reviewed
here.
Other papers related to these subjects and issues are
\cite{iopp}-\cite{Kim:2002gi}.

A first aim of this letter is to study extensions of the above 
improved superconformal type IIA theory which remain exact string theories
in parallel with the analysis performed in \cite{bema} for the type IIB case.
Actually, a partial extension, linked by explicit T-duality to a 
restricted (2,2)
preserving case, was already considered in \cite{iopp}. Here we will give a
notably wider set of possibilities.
Moreover, we will find another class of exact type IIB backgrounds
by using a different manifest superconformal picture.
This will turn out to be the natural picture to describe Type IIB strings 
in pp-wave backgrounds with non constant RR 3-form curvature.

\section{Type IIA exact strings in non constant RR 2 and 4-forms}

\subsection{The Type IIA $U(3)\times U(1)$ formalism}

Because of the different chirality assignment between left and right moving 
fermions, as it was shown in \cite{iopp,leuven},
the analogous type IIA (2,2) $\sigma$-model has three
chiral superfields $\Phi^{\bar i}$,
where $i=1,2,3$, and a twisted chiral one that we call $\Sigma$.
This is the type IIA counterpart of the above superfields formalism in which
only a $U(3)\times U(1)$ symmetry is manifest because of the
necessity of introducing an asymmetric chirality structure to the four supermultiplets.
The type IIA action is then given by
\be
S_0^{IIA}=\frac{1}{2}
\int d^2z d^4\kappa \left\{-\Sigma\bar\Sigma+ \Phi^i\Phi^{\bar i}
\right\} 
\l{22GSA}\ee
The twisted chiral superfields satisfy by definition
$D_-\Sigma=0$, $\bar D_+\Sigma=0$
and
$\bar D_-\bar\Sigma=0$, $D_+\bar\Sigma=0$.

The action (\ref{22GSA}) can be obtained from (\ref{22GSB}) by
T-dualizing along one of the eight transverse directions, and vice versa.
More explicitly\footnote{We follow and shortly review the approach of
  \cite{ghr} and \cite{quotients}.}
let $X$ and $\bar X$ be chiral 
and anti-chiral super fields respectively,
$B$ a general real superfield,
$\Sigma$ and $\bar\Sigma$ twisted chiral and anti-chiral super fields.
Let us consider the action
\be
S_0=\int d^2z d^4\kappa\left(\frac{1}{4}B^2 -\frac{1}{2} B(\Sigma+\bar\Sigma)\right)
\l{free}\ee
We can consider the theory defined by the action $S_0$ in two equivalent ways.

We can complete the squares
$$S_0=\int d^2z d^4\kappa
\left(\frac{1}{4}(B-\Sigma-\bar\Sigma)^2 - \frac{1}{4}(\Sigma+\bar\Sigma)^2 
\right)
$$
and integrate over the decoupled field $B'=B-\Sigma-\bar\Sigma$
by leaving an effective action 
$$\tilde S_0(\Sigma)=\int d^2z d^4\kappa\frac{-1}{4}(\Sigma+\bar\Sigma)^2 =
-\frac{1}{2}\int d^2z d^4\kappa \Sigma\bar\Sigma
$$
Otherwise, we can integrate over 
$\Sigma$ and $\bar\Sigma$
which induce the constrains 
$$\bar D_+D_-B=0 \quad {\rm and} \quad D_+\bar D_-B=0$$
These constrains are solved by
$B=X+ \bar X$ for some chiral $X$ (and anti-chiral $\bar X$).
Recasting this in the remnant of $S_0$, we find the
dual description 
$$S_0(X)=\int d^2z d^4\kappa\frac{1}{4}\left(X+ \bar X\right)^2
=\frac{1}{2}\int d^2z d^4\kappa X\bar X
$$

The same completing argument which applied to 
(\ref{22GSB}), can be invoked for (\ref{22GSA}) too.
In detail, following \cite{bema}, we add four chiral fermionic multiplets
$\Theta^\pm$ and $\bar\Theta^\pm$ and 
four semichiral fermionic multiplets $W^\pm$ and $\bar W^\pm$
and
\footnote{In the following we will refer to this sector of the theory as to the $\Theta W$
  sector.}
improve the action (\ref{22GSA}) by the addition of
\be
S_{\theta w}=\frac{1}{2}\int d^2z d^4\kappa \left\{W^+\Theta^- + W^-\Theta^+ +
\bar W^+\bar\Theta^- + \bar W^-\bar\Theta^+\right\}
\l{thetaw}\ee
where
$ \bar D_\pm W^\mp=0$, $D_\pm\bar W^\mp=0$ 
are the conditions defining the semichiral super-fields
and
$D_\pm\Theta^\mp=0$, 
$D_\pm\bar\Theta^\mp=0$
$\bar D_\pm\Theta^\mp=0$
$\bar D_\pm\bar\Theta^\mp=0$
are the conditions defining the chiral super-fields.
Since the central charges of chiral and twisted chiral multiplets 
are of course equals, the improved theory is manifestly superconformal
in the type IIA case too.

Notice that the improved action 
$S_0^{IIA} + S_{\theta w}$
is invariant under the following supersymmetry transformations
\be
\delta\Theta^\pm=\ep^\pm
\qquad
\delta\bar\Theta^\pm=\bar\ep^\pm
\l{susy1}\ee
\be
\delta\Phi^{i}=\ep^{+i}\Theta^- + \bar\ep^{+i}\bar\Theta^-
, \quad
\delta\Phi^{\bar i}=\ep^{-\bar i}\Theta^+ + \bar\ep^{-\bar i}\bar\Theta^+
\l{susy2}\ee
\be
\delta\Sigma=0
, \quad
\delta\bar\Sigma=0
\l{susy3}\ee
\be
\delta W^+=-\epsilon^{+i}\Phi^{\bar i} 
, \quad
\delta W^-=-\epsilon^{-\bar i}\Phi^{ i}
\l{susy4}\ee
\be
\delta \bar W^+=-\bar\epsilon^{+i}\Phi^{\bar i}
, \quad
\delta \bar W^-=-\bar\epsilon^{-\bar i}\Phi^{ i}
\l{susy5}\ee
as well as under the fermionic symmetries
\be
\delta'\Sigma=\bar D_+ \bar W^+ \bar\zeta + D_- W^-\eta
, \quad
\delta'\bar \Sigma=D_+ W^+\zeta + \bar D_- \bar W^-\bar\eta
\l{ferm1}\ee
\be
\delta'\Theta^+=D_-\bar\Sigma\eta
, \quad
\delta'\Theta^-=D_+\Sigma\zeta
\l{ferm11}\ee
\be
\delta'\bar\Theta^+=\bar D_-\Sigma\bar\eta
, \quad
\delta'\bar\Theta^-=\bar D_+\bar\Sigma\bar\zeta
\l{ferm2}\ee
\be
\delta' W^\pm=0
, \quad
\delta' \bar W^\pm=0
, \quad
\delta' \Phi^i=0
, \quad
\delta' \Phi^{\bar i}=0
\l{ferm3}\ee
which have no direct counterpart in the manifest $U(4)$ Type IIB case
(their Type IIB counterpart will appear in the next section).

\subsection{pp-waves in type IIA}

Due to the light-cone interpretation of the $\Theta W$ sector, looking for 
pp-wave backgrounds corresponds to add to the free improved action
$ S_0^{IIA} + S_{\theta w} $
a further interacting term which does not 
modify the $\Theta$'s equation of motion. Equivalently, requiring $x^-$
independence of the background, implies the independence 
of any further interaction term upon the $W$'s.

The general type IIA supergravity background we refer to is given by the
metric
\footnote{We already distinguish the two directions along which the scalar
  component of the twisted multiplet elongates the string and the six
  direction relative to the chiral triplet superfield scalar components.}
\be
ds_{10}^2=-2dx^+dx^- +H(dx^+)^2 + d\phi^i d\phi^{\bar i} +d\sigma d\bar\sigma
\l{bg1}\ee
and the RR-field curvatures
\be
F_{(4)}=dx^+\wedge\omega_{(3)}\qquad 
F_{(2)}=dx^+\wedge\omega_{(1)}
\l{bg2}\ee
This class of backgrounds satisfies the supergravity equations of motion 
if the $\omega_{(i-1)}$ are harmonic forms in the transverse eight dimensional space
and 
if the pp-wave factor $H$ satisfies the equation 
\be
\left(\partial_i\partial_{\bar i}+\partial_\sigma\partial_{\bar\sigma}\right)H
\propto
|\omega_{(3)}|^2+|\omega_{(1)}|^2
\l{eom}\ee
which is the only non trivial equation for the metric, namely $R_{++}+T_{++}=0$.

Therefore, we consider the most general classically (2,2) superconformal
action for pp-waves that is
\be
S_{pp}=S_0^{IIA} + S_{\theta w} + \int d^2z d^4\kappa
U\left(\Phi^{i},\Sigma,\Phi^{\bar i},\bar\Sigma,\Theta^\pm,\bar\Theta^\pm\right)
\l{ppgen}\ee
where $U$ is a general bosonic superfield function.
Restricting to polynomial dependence on the superfields and requiring 
that $U$ effectively generates RR vertices, we consider
a general bosonic superfield at most quadratic in the $\Theta$s and containing an 
odd number of barred and unbarred fermions.

We would like now to specify few cases preserving some of the
supersymmetries (\ref{susy1}-\ref{susy5}) above and which are 
exactly superconformal.

\paragraph{(2,2) supersymmetry}

A first class is the one with all (2,2) supersymmetry,
i.e. the $\ep^\pm$ and the $\bar\ep^\pm$,
where one takes
$$
U\left(\Phi^{i},\Sigma,\Phi^{\bar i},\bar\Sigma,\Theta^\pm,\bar\Theta^\pm\right)
= $$ $$=
\Theta^+\bar\Theta^+{\cal F}(\Phi^i)
+
\Theta^-\bar\Theta^-\bar{\cal F}(\Phi^{\bar i})
+
\Theta^+\bar\Theta^-{\cal E}(\Sigma)
+
\Theta^-\bar\Theta^+\bar{\cal E}(\bar \Sigma)
$$
where ${\cal F}$ and ${\cal E}$ are holomorphic functions of 
the $\Phi^i$'s and $\Sigma$ respectively.

Notice that the full action is still invariant under the supersymmetries 
$\ep^\pm$ and $\bar\ep^\pm$
if we modify the $\delta W=0$ rule (with respect to these variations) to 
\be
\delta W^-=\bar\ep^+{\cal F}+\bar\ep^-{\cal E}
\quad
\delta W^+=\bar\ep^-\bar{\cal F}+\bar\ep^+\bar{\cal E}
\l{modiss1}\ee
\be
\delta \bar W^-=-\ep^+{\cal F}-\ep^-\bar{\cal E}
\quad
\delta \bar W^+=-\ep^-\bar{\cal F}-\ep^+{\cal E}
\l{modiss2}\ee
This is consistent with the semichiral properties of the $W$s.

It is also possible to modify the fermionic symmetry (\ref{ferm1}-\ref{ferm3})
in such a way that the action remains invariant. 
This happens if ${\cal F}=0$ and, in this case,
one just has to modify the $\delta' W=0$ rule in (\ref{ferm3}) to 
\be
\delta' W^-=-{\cal E}\bar D_+\bar\Sigma\bar\zeta
\quad
\delta' W^+=-\bar{\cal E}\bar D_-\Sigma\bar\eta
\l{modferm1}\ee
\be
\delta' \bar W^-=-\bar{\cal E}D_+\Sigma\zeta
\quad
\delta' \bar W^+=-{\cal E}D_-\bar\Sigma\eta
\l{modferm2}\ee
Notice that this peculiar extension shows the nature of this further fermionic 
symmetry which acts transversely with respect to the chiral supersymmetry 
structure.
Notice, in particular, that if the chiral superpotential ${\cal F}$ 
is left non vanishing, then one finds
in the $\delta'$ variation of the action terms like 
$
\int \theta^-\bar{\cal F} \bar D_+\bar\Sigma\bar\zeta
$
which can not be reabsorbed nor with a consistent shift in $\delta'\Sigma$
nor in $\delta' W^+$, where consistency is understood with respect to the chiral/anti-chiral
constrains on the superfields.

The same argument as the one given in \cite{bema}, leads to exact conformal
invariance of these models.
Notice that, upon on-shell reduction of the $\Theta W$ sector,
the (2,2) $\sigma$-model action is 
$$
\frac{1}{2}
\int d^2z d^4\kappa \left\{-\Sigma\bar\Sigma+ \Phi^i\Phi^{\bar i}
\right\} 
+
$$
\be
+\frac{1}{2}
\int d^2z d\kappa^-d\bar\kappa^-{\cal F}(\Phi^i)|_{\kappa^+=0=\bar\kappa^+}
+
\int d^2z d\kappa^-d\bar\kappa^+{\cal E}(\Sigma)|_{\kappa^+=0=\bar\kappa^-}
+\quad{h.c.}
\l{2,2}\ee

This corresponds to the following type IIA background.
The pp-wave factor $H$ in (\ref{bg1}) is 
$$
H=|\partial_i{\cal F}|^2 + |\partial_\sigma{\cal E}|^2
$$
and the harmonic forms are
$$
\omega_{(1)}=\sqrt{2} {\rm Re}\left\{\partial_\sigma^2{\cal E} d\sigma\right\}
$$
and
$$
\omega_{(3)}={\rm Re}\left\{\epsilon_{\bar i\bar j}^{\,\,\,
    l}\partial_l\partial_k
{\cal F} d\phi^{\bar i}\wedge d\phi^{\bar j}\wedge d\phi^k
\right\}
$$
which solve equation (\ref{eom}).

Notice that if in particular one restricts to the case ${\cal E}=0$, the
background already considered in \cite{iopp} is recovered as a sub-case.

\paragraph{(1,1) extra supersymmetry}

A second class is the one with extra $(1,1)$ supersymmetry $\ep^+=\ep^-$ and $\bar\ep^+=\bar\ep^-$
where one takes 
$$
U\left(\Phi^{i},\Sigma,\Phi^{\bar i},\bar\Sigma,\Theta^\pm,\bar\Theta^\pm\right)
=
\left(\Theta^+-\Theta^-\right)
\left(\bar\Theta^+-\bar\Theta^-\right)
V\left(\Phi^{i},\Sigma,\Phi^{\bar i},\bar\Sigma\right)
$$
and the condition of exact superconformal invariance takes the form of
\be
\left(
\partial_\Sigma\partial_{\bar\Sigma} +
\partial_{i}\partial_{\bar i} 
\right)V=0
\l{11sc}\ee
that is $V$ is an harmonic function in the transverse eight dimensional space.
Notice that the sign in front of the $\partial_\Sigma\partial_{\bar\Sigma}$
in equation (\ref{11sc})
combines the minus sign of the super-propagator of twisted chiral fields 
and the minus sign in the mixed $\Theta^\pm\bar\Theta^\mp$ on shell contractions.
This crucial sign (for comparison with supergravity) can also be checked by 
examining the highest component of the stress-energy tensor and by requiring 
the absence of double pole with $V$. This condition reproduces directly the 
harmonic equation (\ref{11sc}).

This corresponds to the following type IIA background.
The pp-wave factor $H$ in (\ref{bg1}) is 
$$
H=|\partial_\sigma V|^2+|\partial_i V|^2
$$
and the harmonic forms are
$$
\omega_{(1)}= d\left[\partial_\sigma V\right]\quad+\quad{\rm h.c.} 
$$
and
$$
\omega_{(3)}=d\left[\partial_i V\ep^i_{\,\, \bar k\bar l}d\phi^{\bar k}\wedge
d\phi^{\bar l}\quad+\quad{\rm h.c.}
\right]
$$
where $d$ is the usual differential of forms in the transverse eight dimensional space.
Notice that if $V$ is independent on the twisted chiral superfields,
then the above solution could have been obtained by T-duality from
an analogous Type IIB case with $(1,1)$ supersymmetry.

\vspace{.5 cm}

Also for the more general type IIA background geometries that we considered
here it is possible to formulate a matrix string model by the appropriate 
non-abelian gauging of the $\sigma$-model, as in 
\cite{iopp,leuven,ppmst}.

\section{Exact backgrounds for Type IIB with RR 3-form}

\subsection{The $U(2)\times U(2)$ formalism for Type IIB}

As it appears clearly from the introduction,
there exists different possible $(2,2)$ manifest superconformal
formalisms for type II string theories on flat background.
These pictures can be obtained by formally T-dualizing along the way we 
exposed in the introduction.

As far as the type IIA case is concerned, the other possible picture, that is 
with one chiral and three twisted chiral superfields, is equivalent to the one with
three chiral and one twisted chiral superfields 
just by a relabeling of the superspace variables (who is chiral and who is antichiral
makes sense only relatively).
Therefore, the Type IIA picture we studied so far is exhaustive of the inequivalent cases
also after the addition of the RR-background fields.

As far as the type IIB case is concerned, 
we can represent type IIB on a flat background in two equivalent ways.
One is the $U(4)$ manifest picture with four chiral multiplets.
Another one is with two chiral and two twisted chiral multiplets
with action
\be
S'=\frac{1}{2}\int d^2z d^4\kappa 
\left(
\Phi^a\Phi^{\bar a}-\Sigma^{a'}\Sigma^{\bar a'}
\right)
\l{prime}\ee
which exhibits a manifest $U(2)\times U(2)$ rotational symmetry (here $a,a'=1,2$).
The picture (\ref{prime}) is equivalent to the $U(4)$ manifest one (by two 
T-dualities) on the flat background.

Obviously, a picture with four twisted chiral superfields and no chiral one is equivalent 
to the one with four chiral superfields just by a relabeling of 
the superspace variables. 

After the introduction of the RR curvature field backgrounds, the picture (\ref{prime})
is no longer generically  
equivalent to the manifest $U(4)$ and allows the description of other
exact string theories on curved backgrounds of pp-wave type.

Possible inequivalent RR-backgrounds for type IIB 
which preserve some supersymmetry
and are exactly superconformal can be again obtained by the method 
of adding a vertex quadratic in the $\Theta$s which does not generate 
conformal anomalies because of algebraic reasons.

Before studying these backgrounds, we notice that the $\Theta W$
improved type IIB action $S'+S_{\theta w}$
is invariant under the following supersymmetries which act transversely 
with respect to the chiral structure of the $\sigma$-model picture.
These are
\be
\delta\Theta^\pm=\ep^\pm
\qquad
\delta\bar\Theta^\pm=\bar\ep^\pm
\l{susy1'}\ee
\be
\delta\Phi^{a}=\ep^{a}\Theta^- + \bar\ep^{a}\bar\Theta^-
, \quad
\delta\Phi^{\bar a}=\ep^{\bar a}\Theta^+ + \bar\ep^{\bar a}\bar\Theta^+
\l{susy2'}\ee
\be
\delta\Sigma^{a'}=0
, \quad
\delta\Sigma^{\bar a'}=0
\l{susy3'}\ee
\be
\delta W^+=-\epsilon^{a}\phi^{\bar a} 
, \quad
\delta W^-=-\epsilon^{\bar a}\phi^{a}
\l{susy4'}\ee
\be
\delta \bar W^+=-\bar\epsilon^{a}\phi^{\bar a}
, \quad
\delta \bar W^-=-\bar\epsilon^{\bar a}\phi^{a}
\l{susy5'}\ee
as well as under the fermionic symmetries
\be
\delta'\Sigma^{a'}=\bar D_+ \bar W^+ \bar\zeta^{a'} + D_- W^-\eta^{a'}
, \quad
\delta'\Sigma^{\bar a'}=D_+ W^+\zeta^{\bar a'} + \bar D_- \bar W^-\bar\eta^{\bar a'}
\l{ferm1'}\ee
\be
\delta'\Theta^+=D_-\Sigma^{\bar a'}\eta^{a'}
, \quad
\delta'\Theta^-=D_+\Sigma^{a'}\zeta^{\bar a'}
\l{ferm11'}\ee
\be
\delta'\bar\Theta^+=\bar D_-\Sigma^{a'}\bar\eta^{\bar a'}
, \quad
\delta'\bar\Theta^-=\bar D_+\Sigma^{\bar a'}\bar\zeta^{a'}
\l{ferm2'}\ee
\be
\delta' W^\pm=0
, \quad
\delta' \bar W^\pm=0
, \quad
\delta' \Phi^a=0
, \quad
\delta' \Phi^{\bar a}=0
\l{ferm3'}\ee
These are the analogous to Eq. (\ref{ferm1}-\ref{ferm3})
for type IIB in the picture with two chiral and two twisted chiral superfields.

\subsection{Type IIB pp-waves with RR 3-form backgrounds}

Also the picture (\ref{prime}) allows a family of interaction vertices 
preserving some supersymmetries and manifest conformal invariance.

These parallel the (1,1) and (2,2) preserving cases given for type IIA
in the previous section and for type IIB in \cite{bema}.
The novelty in the case at hand is that the pp-waves 
naturally support 3-form RR field curvatures.

The background we consider now is
\be
ds_{10}^2=-2dx^+dx^- +H(dx^+)^2 + d\phi^a d\phi^{\bar a} +d\sigma^{a'} d\bar\sigma^{\bar a'}
\l{bg1'}\ee
and the RR-field curvature
\be
F_{(3)}=dx^+\wedge\omega_{(2)}\qquad 
\l{bg2'}\ee
This class of backgrounds satisfies the supergravity equations of motion 
if $\omega_{(2)}$ is harmonic in the transverse eight dimensional space
and 
if the pp-wave factor $H$ satisfies the equation 
\be
\left(\partial_a\partial_{\bar a}+\partial_{a'}\partial_{\bar a'}\right)H
\propto
|\omega_{(2)}|^2
\l{eomprime}\ee
which is the non trivial equation for the metric.

\paragraph{(1,1) preserving case}

The {\it (1,1) preserving case} corresponds to the addition to the $\Theta W$-improved 
action 
$S'+S_{\theta w}$
of the vertex 
$$
U\left(\Phi^{a},\Sigma^{a'},\Phi^{\bar a},\Sigma^{\bar a'},\Theta^\pm,\bar\Theta^\pm\right)
=
\left(\Theta^+-\Theta^-\right)
\left(\bar\Theta^+-\bar\Theta^-\right)
V\left(\Phi^{a},\Sigma^{a'},\Phi^{\bar a},\Sigma^{\bar a'}\right)
$$
and the condition of exact superconformal invariance takes the form of
\be
\left(
\partial_a\partial_{\bar a} +
\partial_{a'}\partial_{\bar a'} 
\right)V=0
\l{11sc'}\ee
that is $V$ again a harmonic function in the transverse eight dimensional space.

This corresponds to the following type IIB background.
The pp-wave factor $H$ in (\ref{bg1'}) is 
$$
H=|\partial_a V|^2+|\partial_{a'} V|^2
$$
and the harmonic 2-form is
$$
\omega_{(2)}= d\left[\partial_a V \ep^a_{\bar b}d\phi^{\bar b}+
\partial_{a'} V \ep^{a'}_{\bar b'}d\phi^{\bar b'}
\right]\quad+\quad{\rm h.c.} 
$$
where $d$ is the form differentiation in the transverse eight dimensional space.

\paragraph{(2,2) preserving case}

The {\it (2,2) preserving case} corresponds instead to the addition to the $\Theta W$-improved 
action of the vertex 
$$
U\left(\Phi^{a},\Sigma^{a'},\Phi^{\bar a},\Sigma^{\bar a'},\Theta^\pm,\bar\Theta^\pm\right)
=$$ $$=
\Theta^+\bar\Theta^+{\cal F}(\Phi^a)
+
\Theta^-\bar\Theta^-\bar{\cal F}(\Phi^{\bar a})
+
\Theta^+\bar\Theta^-{\cal E}(\Sigma^{ a'})
+
\Theta^-\bar\Theta^+\bar{\cal E}(\Sigma^{\bar a'})
$$
where ${\cal F}$ and ${\cal E}$ are holomorphic functions of 
the $\phi^a$'s and $\Sigma^{a'}$ respectively.

Notice that the full action is still invariant under the supersymmetries 
$\ep^\pm$ and $\bar\ep^\pm$
if we modify the $\delta W=0$ rule (with respect to these variations) 
as for the type IIA case to 
\be
\delta W^-=\bar\ep^+{\cal F}+\bar\ep^-{\cal E}
\quad
\delta W^+=\bar\ep^-\bar{\cal F}+\bar\ep^+\bar{\cal E}
\l{modiss1'}\ee
\be
\delta \bar W^-=-\ep^+{\cal F}-\ep^-\bar{\cal E}
\quad
\delta \bar W^+=-\ep^-\bar{\cal F}-\ep^+{\cal E}
\l{modiss2'}\ee
Again this is consistent with the semichiral properties of the $W$s.

It is also possible to modify the fermionic symmetry (\ref{ferm1'}-\ref{ferm3'})
in such a way that the action remains invariant.
This happens if ${\cal F}=0$ and, in this case,
one just has to modify the $\delta' W=0$ rule in (\ref{ferm3'}) to 
\be
\delta' W^-=-{\cal E}\bar D_+\Sigma^{\bar a'}\bar\zeta^{a'}
\quad
\delta' W^+=-\bar{\cal E}\bar D_-\Sigma^{a'}\bar\eta^{\bar a'}
\l{modferm1'}\ee
\be
\delta' \bar W^-=-\bar{\cal E}D_+\Sigma^{a'}\zeta^{\bar a'}
\quad
\delta' \bar W^+=-{\cal E}D_-\Sigma^{\bar a'}\eta^{a'}
\l{modferm2'}\ee
Notice that this peculiar extension shows again the nature of this further fermionic 
symmetry which acts transversely with respect to the chiral supersymmetry 
structure.

\vspace{.5 cm}

Let us notice that the construction performed in this section  is 
complementary to the analysis of these backgrounds done with other 
methods by Russo and Tseytlin in \cite{ruts}.

\section{Conclusions and Open Questions}

In this letter we developed and generalized 
the approach first proposed by Berkovits and Maldacena
in \cite{bema}
to characterize exact string theories on pp-wave backgrounds sustained
by non constant RR curvature forms.
Our main two results are the generalization of the hybrid formalism
to type IIA string theory and the formulation of a complementary
hybrid formalism for type IIB.
By using these tools, we are able to describe new exact superconformal
$\sigma$-models for
string theories coupled to new pp-wave backgrounds sustained by different
RR curvature forms.

\subsection{Other backgrounds}

Another possible class of backgrounds which can be shown 
easily to be exact are the ones with constant 
$B$ field in the eight dimensional transverse space.

That's because, as shown in \cite{ghr}, mixed terms in the quadratic free 
action
involving chiral and twisted chiral correspond to constant $B$ fields.
Therefore, we can add for free to (\ref{22GSA}) the term
\footnote{There is an analogous one for (\ref{prime}) too.}
\be
\int d^2z d^4\kappa 
\left[B^1_i\phi^i +B^2_{\bar i}\phi^{\bar i}\right] \Sigma +
\quad{\rm h. c.}
\l{cbf}\ee
Since the $\sigma$-model action is still free, it is trivially 
superconformal exact.
This corresponds to a flat ten dimensional background with a constant $B$
field turned on in the eight directions transverse to the light-cone.

Yet another class of backgrounds which are exactly superconformal
are the dilatonic ones studied in \cite{Berkovits:2002rd}. These
can be adapted easily to the type IIA string too.

A more general class of Type IIA backgrounds 
which it would be very nice to understand in the superfield 
formalism is that of pp-waves
warped by an harmonic functions as
\be
ds_{10}^2= A^{-3/4}\left[
-2dx^+dx^- +H(dx^+)^2 \right]
+ 
A^{1/4}\left[
d\phi^i d\phi^{\bar i} +d\sigma d\bar\sigma\right]
\l{bg'1}\ee
with a dilaton and NSNS-field curvature 
\be
\Phi=-\frac{1}{2}{\rm log}A
\qquad
G_{(3)}=dA^{-1}\wedge dx^+\wedge dx^-
\l{bg'2}\ee
and the RR-field curvatures
\be
F_{(4)}=dx^+\wedge\omega_{(3)}\qquad 
F_{(2)}=dx^+\wedge\omega_{(1)}
\l{bg'3}\ee
This class of backgrounds satisfies the supergravity equations of motion 
if the $\omega_{(i-1)}$ are harmonic forms in the transverse eight dimensional space,
if the pp-wave factor $H$ satisfies the equation 
\be
\left(\partial_i\partial_{\bar i}+\partial_\sigma\partial_{\bar\sigma}\right)H
\propto
|\omega_{(3)}|^2+|\omega_{(1)}|^2
\l{eom'}\ee
and if the warp factor $A$ is an harmonic function in the eight dimensional space.

These generalize the backgrounds studied in \cite{clps},
naively reduce for $A=1$ to the ones we studied in 
Section 2 of this letter
and are one of their most natural generalizations.
A possibility for their implementation in the above formalism \cite{bpc}
maybe in a non trivial $W$ dependence in the improved $\sigma$-model action
in the style of \cite{Berkovits:2002rd} to reproduce the non constant
warping factor and dilaton.

\subsection{Non-local symmetries and RR-vertices}

There is increasing evidence that the M-theory picture for string theory has to
include also bosonic string.
This has been discussed for example in \cite{LE}, where 
ten dimensional superstrings on flat space-time, as well as Type 0s, have been rebuilt 
from the critical bosonic 26 dimensional string theory in flat spacetime.
On the other way around, the embedding of $N=i$ in $N=i+1$, for $i=0,1$,
string theories was established in \cite{Berkovits:1993xq}.
If it is true that all known string theories are one,
it has to be true that for any consistent choice of the background,
superstring has a purely bosonic string counterpart on some related 
dual background geometry.

It is therefore important to understand how this happens, or has chances
to happen, on the exact backgrounds that we are considering.
A first suggestion has been given in \cite{Gomez}, where 
the $N=0$ to $N=2$ correspondence for certain integrable 
systems has been recalled from \cite{LeClair:1992bj}.

In summary, the correspondence is a duality between the chiral ring structure 
of a $N=2$ theory, which is defined by its superpotential function, and the
vertex algebra in the integrable $N=0$ theories.
Algebraically, the map is established at the points in the moduli space of 
couplings of $N=0$ theories where a ${\cal U}_q\left(\hat sl(2)\right)$ 
affine quantum symmetry arises with $q=\pm i$. 
In this case, the affine quantum algebra structure reproduces directly
the $N=2$ superalgebra
\footnote{It is very much tempting to consider the more general Drinfeld 
quantum double of the Yangian ${\cal U}_q\left(\hat {\tt g}\right)$
symmetry of massive WZNW models -- i.e. such that the kinetic term 
coefficient is not matched to the level ${\tt k}$ which appears as prefactor 
of the WZ term -- as the $N=0$ starting point for a generalization.}.

Typically, to reproduce the kind of background we are considering here, namely
with non compact transverse eight dimensional space -- due to
the fact that we require the existence of non constant holomorphic functions
on such a space --
one might consider generalized (non conformal) affine Toda theories.

The full string theory off shell symmetry, namely the conformal symmetry,
might then be recovered via further affinization (as suggested in \cite{Gomez}
for the simpler case), by generalizing the construction \cite{bb}.
It remains a crucial issue to understand if the $\Theta W$ sector completition 
of the $N=2$ theory is map to the affinization to the conformal theory of the 
$N=0$ model counterpart or if this point is still problematic
and requires a deeper analysis.

Notice that there is a further link between the
quantum groups invariance and a particular feature
of string theory on RR-background.

As it is well known, the quantum group symmetry 
which is typical of the massive integrable two dimensional
field theories we are using to model string theories in RR-background
geometries,
extends non trivially
the classical Colemann-Mandula theorem because it does not act on
multiparticle states as if they were tensor products of one particle states.
Technically, this happens because of the non triviality of the
comultiplication as an Hopf algebra.
Equivalently, from the quantum field theory point of view,
the preserved charges are given by integrals of 
time components of currents which are not local with respect to the 
elementary fields creating one particle states.
Therefore, the symmetry is of a non local type and requires to be 
realized in a theory with vertices which are non local with respect to the 
currents.

On the string theory side, we know that the equivalence of the GS formalism 
to the covariant RNS is non local, because of the bosonization involved while
implementing the $SO(8)$ triality transformations for the 
world sheet fermion fields. 
As a consequence, the RR-vertices are mapped to bilinears of the spin 
operators which are non local with respect to the world-sheet fields. 
This generates 
well known problems in forcing the RNS description for RR-backgrounds
with the usual local $\sigma$-model techniques.

We believe that these manifestation of non-localities in the 
two dimensional quantum field theory algebras should be somehow linked
and deeping this analysis could hopefully clarify the intricacies
of the nonlocal change of variable \cite{Berkovits:2002rd}
needed to link our approach to the covariant R-NS one.

\vspace{.5 cm}

In conclusion let us add still a couple of observations. 
The extension to non flat transverse eight-dimensional space geometries
of the formalisms which has been developed here is possible.
Moreover, it would be of extreme interest to have a link with 
the techniques of \cite{notation,mirror} to describe 
string theories on Calabi-Yau's and geometric transitions 
in presence of RR-backgrounds at the microscopic $\sigma$-model level.

\ack
I would like to thank N. Berkovits, M. Bertolini, C. Gomez, A. Hammou, 
M. Henneaux, N. Kim, R. Russo and A. Sevrin
for discussions and e-mail exchange.
In particular I would like to thank N. Berkovits for reading the manuscript 
and for very useful criticisms.
\noindent
This work is supported by the Marie Curie fellowship contract
HPMF-CT-2002-0185

\end{document}